# Performance analysis of a novel hybrid FSO / RF communication system

*Mohammad Ali Amirabadi[1]*
[1] School of Electrical Engineering, Iran University of Science and Technology, Tehran, Iran
✉ E-mail: m_amirabadi@elec.iust.ac.ir

**Abstract:** In this paper, a novel dual-hop relay-assisted hybrid Free Space Optical / Radio Frequency (FSO / RF) communication system is presented. In this structure an access point connects users within the building to the Base Station via a hybrid parallel FSO / RF link, this link is proposed firstly. Parallel combination of FSO and RF links and use of an access point, will increase capacity, reliability and data rate of the system. It is the first time that the effect of number of users on the performance of a dual-hop relay-assisted hybrid parallel FSO / RF system is investigated. FSO link is considered in Gamma-Gamma atmospheric turbulence with the effect of pointing error and RF link is considered in Rayleigh fading. For the first time, closed-form expressions are derived for Bit Error Rate (BER) and Outage Probability ($P_{out}$) of the proposed system. Derived expressions are verified through MATLAB simulations. It is shown that the performance of the proposed system is almost independent of atmospheric turbulence intensity, thereby when atmospheric turbulence strengthens, low power consumption is required for maintenance of the system performance. Hence the proposed structure is particularly suitable for mobile communication systems in which a small mobile battery supplies transmitter power. Also the proposed system performance of the system is preferable even at low signal to noise ratio (SNR). Therefore, proposed structure significantly reduces power consumption while maintaining performance of the system.

## 1 Introduction

According to the data traffic increase in modern communication systems, it is predicted that the 5th generation cellular networks, in comparison with the 4th generation, requires 1,000 times more capacity, 10 times more spectral (energy) efficiency, 10-100 times more data rate, and 25 times more average cell throughput. Ultra-dense cellular network, with a large number of small cells, is one of the proposed solutions to increase the capacity. The data traffic of small cells is transmitted to the core through a backup network. Combination of optical fibre link and millimetre-wave RF link is a good candidate for such a backup network. However, in ultra-dense cellular networks, RF interference is a problematic issue and also capacity of millimetre-wave RF link is not enough for the requirements of the 5th generation cellular network. In addition, optical fibre has high installation cost. FSO link has bandwidth and data rate equal to optical fibre link, therefore, in hybrid structures, it is better to replace the optical fibre by FSO link [1].

FSO link is sensitive to atmospheric turbulence and weather conditions. Also the effect of pointing error, which is caused by transceiver misalignment, significantly degrades performance of the FSO system. Weather conditions apply constant loss on the received signal intensity, but atmospheric turbulences and pointing error lead to random fluctuations of the received signal intensity [2].

In order to investigate the effects of atmospheric turbulences, various statistical models have been presented in papers. Among them, Log-normal, Gamma-Gamma and Negative Exponential models are respectively in high compliance with experimental results, for weak, moderate-to-strong, and saturate atmospheric turbulence regimes. Mostly, in FSO system, Intensity Modulation / Direct Detection (IM/DD) is used, among coherent detections, heterodyne detection is a complex technique that overcomes the effect of thermal noise [3].

FSO system has unlicensed bandwidth, inherent security and easy setup. Combining FSO and millimetre-wave RF links significantly increases data rate and reliability of the system. Works done about hybrid FSO / RF systems can be divided in three main categories. The first category has one hop structure [4-8], in which link reliability and data rate are significantly increased by implementation of a parallel FSO / RF link. In this structure either both FSO and RF links are always active or FSO link is always active and RF link acts as a backup [9]. Second category investigates performance of dual-hop structures [10-13]. This category uses a relay which improves capacity and reduces total power consumption of the system. Several protocols have been proposed for data processing in relay-assisted hybrid FSO / RF systems, among them amplify and forward [14] and detect and forward [15] are mostly used. In amplify and forward protocol, amplification gain is fixed or adaptive. Fixed gain has less complexity but more power dissipation, thereby it is better be used only when CSI is unknown. When CSI is known, detection and forward, due to its low power consumption is preferred. Third category deals with multi hop structures [16-20]. These structures reduce total power consumption and increases throughput of the system.

Many works with different structures have been done about single and dual hop hybrid FSO/RF communication system. Papers with Single-hop, have implemented parallel FSO/RF link [5-8], some of them used the idea of multi-user [4]. A new paper has investigated use of diversity in this area for the first time [21]. Dual hop papers mostly used series FSO and RF structure with RF at the first and FSO at the second hop [10-15, 22-24]. Sometimes a FSO or RF link is also available as a backup between source and destination nodes [25, 26]. According to the best of the authors' knowledge, this is the first time that a parallel FSO/RF link is implemented in a dual-hop hybrid FSO/RF system. Use of parallel FSO/RF link significantly improves performance and reliability of the system. Because there is almost no atmospheric condition which could degrade performance of both FSO and RF links at the same time. In this paper, for the first time, the performance of a dual-hop hybrid FSO / RF system is investigated from point of view of number of users, and it is the first time that opportunistic transmission scheme on multiuser RF/FSO system is deployed. This significantly improves proposed system performance. Also dual-hop relaying has the advantages of increasing system capacity as well as decreasing total power consumption.

In this work, FSO link is considered at Gamma-Gamma atmospheric turbulence because this model is highly accompanied with experimental results. Also in order to get closer to actual operations, the effect of pointing error is considered at FSO link. RF link is at Rayleigh fading. In this paper, two cases of known CSI and unknown CSI at the access point are considered. In the case of known CSI, received signal at the access point is detected, regenerated and then forwarded, In the case of unknown CSI, received signal at the access point is amplified with fixed gain and forwarded. For the first time, closed-form expressions are derived for BER and $P_{out}$ of the proposed structure. Hybrid parallel FSO/RF structure significantly improves link accessibility and reliability as well as data rate, also the use of an access point improves capacity of the system; hence the proposed structure will reduce power consumption while maintaining performance of the system.

Rest of the paper is organized as follows: in section 2 system model is described. Sections 3 and 4 investigate performance of known CSI and unknown CSI schemes, respectively. Section 5 compares simulation and analytical results and brings some discussions. Section 6 is the conclusion of this work.

## 2 System Model

The dual-hop hybrid FSO / RF communication system of Fig. 1 is considered, where mobile users communicate with the destination via the intermediate access point by adopting amplify and forward or detect and forward relaying schemes.

Let $x_{1,i}; i = 1,2,...,N$ be the transmitted RF signal from $i-th$ user to access point through RF link. The received signal at the access point is expressed as:

$$y_{1,i} = h_i x_{1,i} + e_{1,i}, \quad (1)$$

where $h_i$ represents the fading coefficient of the RF link between $i-th$ user and the access point, and $e_{1,i}$ represents the additive white Gaussian noise (AWGN) at access point input with variance of $\sigma_{RF}^2$ and zero mean.

Received signal with the highest SNR is selected at the access point. Instantaneous SNR at the access point input is written as:

$$\gamma_1 = \max(\gamma_{1,1}, \gamma_{1,2}, ..., \gamma_{1,N}). \quad (2)$$

In this system, in the case of known CSI, the received RF signal with the highest SNR at the access point input is detected, regenerated, then two copies of the generated signal are forwarded through parallel FSO and RF links. At the FSO link, RF signal is first converted to FSO by conversion efficiency of $\eta$, then a direct current (DC) bias is added to the ensure that the FSO signal is non-negative. Transmitted FSO and RF signals are as follows:

$$x_{2,FSO} = 1 + \eta d_1, \quad (3)$$
$$x_{2,RF} = d_1, \quad (4)$$

where $d_1$ is the regenerated signal. Transmitted signal is affected by channel atmospheric turbulence and receiver input noise. Finally, after DC removal from FSO signal, received FSO and RF signals at the Base Station are given by:

$$y_{2,FSO} = I_2 x_{2,FSO} + e_{2,FSO} - I_2 = I_2 \eta d_1 + e_{2,FSO}, \quad (5)$$
$$y_{2,RF} = h_2 x_{2,RF} + e_{2,RF} = h_2 d_1 + e_{2,RF}, \quad (6)$$

where $I_2$ is atmospheric turbulence intensity along FSO link, $e_{2,FSO}$ represents the AWGN with zero mean and variance of $\sigma_{FSO}^2$ at the FSO receiver input, $h_2$ is RF fading coefficient, and $e_{2,RF}$ is AWGN with zero mean and variance of $\sigma_{RF}^2$ at the RF receiver input.

In the case of unknown CSI, the received RF signal with the highest SNR at the access point input is selected, amplified with fixed gain, then two copies of it are forwarded through parallel FSO and RF links as:

$$x_{2,FSO} = G(1 + \eta y_1), \quad (7)$$

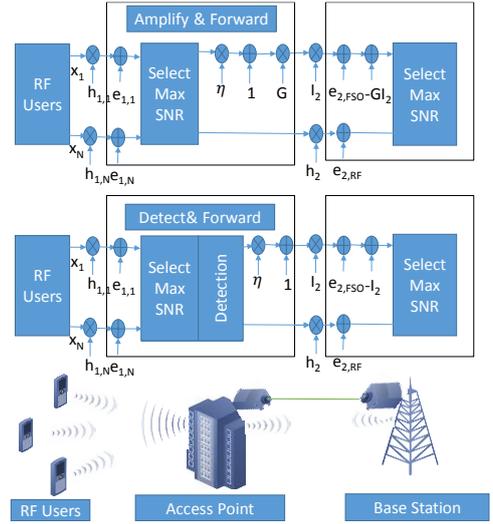

**Fig. 1**: The proposed relay-assisted hybrid FSO / RF system.

$$x_{2,RF} = G y_1, \quad (8)$$

where $y_1$ is the selected signal. Received FSO and RF signals at the Base Station are given by:

$$y_{2,FSO} = I_2 x_{2,FSO} + e_{2,FSO} - GI_2 = G\eta I_2 y_1 + e_{2,FSO}, \quad (9)$$
$$y_{2,RF} = h_2 x_{2,RF} + e_{2,RF} = G h_2 y_1 + e_{2,RF}. \quad (10)$$

Assuming transmitted signal with unit energy, the instantaneous SNRs at Base Station receivers input are as follows:

$$\gamma_{FSO} = \frac{G^2 \eta^2 I_2^2 h_1^2}{G^2 \eta^2 I_2^2 \sigma_{RF}^2 + \sigma_{FSO}^2}, \quad (11)$$

$$\gamma_{RF} = \frac{G^2 \eta^2 h_2^2 h_1^2}{G^2 h_2^2 \sigma_{RF}^2 + \sigma_{RF}^2}. \quad (12)$$

Regarding the fixed-gain relay strategy at the unknown CSI scheme, its gain is fixed to a constant value, which is independent of the CSI of the first-hop channel. The amplification gain is fixed to $G^2 = 1/(C\sigma_{RF}^2)$, where $C$ is a constant parameter. Defining $\gamma_1 = h_1^2/\sigma_{RF}^2$, $\gamma_{2,RF} = G^2 h_2^2/\sigma_{RF}^2$ and $\gamma_{2,FSO} = G^2 \eta^2 I_2^2/\sigma_{FSO}^2$, instantaneous SNR at the Base Station receivers input become as follow:

$$\gamma_{FSO} = \frac{\gamma_1 \gamma_{2,FSO}}{\gamma_{2,FSO} + C}, \quad (13)$$

$$\gamma_{RF} = \frac{\gamma_1 \gamma_{2,RF}}{\gamma_{2,RF} + C}. \quad (14)$$

Since the amplification gain is fixed, the forwarded signal has a varying output power, due to effect of the channel fading by the first hop before their fixed-gain amplification. At the Base Station, between received FSO and RF signals, signal with higher SNR is selected for detection. Therefore instantaneous SNR at the Base Station receiver input is as follows:

$$\gamma_2 = \max(\gamma_{FSO}, \gamma_{RF}). \quad (15)$$

In present study, FSO link is investigated in wide range of atmospheric turbulence regimes, from moderate to strong. The best model that accompany experimental results of this range is Gamma-Gamma distribution. Also the effect of pointing error are considered in order to get closer to actual results. Cumulative Distribution Function (CDF) of Gamma-Gamma atmospheric turbulence with the effect of pointing error is as follows [27]:

$$F_\gamma(\gamma) = \frac{\xi^2}{\Gamma(\alpha)\Gamma(\beta)} G_{2,4}^{3,1}\left(\alpha\beta\kappa\sqrt{\frac{\gamma}{\bar{\gamma}_{FSO}}}\bigg|\begin{matrix}1,\xi^2+1\\ \xi^2,\alpha,\beta,0\end{matrix}\right), \quad (16)$$

where $G_{p,q}^{m,n}\left(z\bigg|\begin{matrix}a_1,a_2,\ldots,a_p\\ b_1,b_2,\ldots,b_q\end{matrix}\right)$, is Meijer-G function [28, Eq. 07.34.02.0001.01]0, $\Gamma(.)$ is well known Gamma function [28, Eq. 06.05.02.0001.01] $\alpha = \left[\exp\left(0.49\sigma_R^2/(1+1.11\sigma_R^{12/5})^{7/6}\right)-1\right]^{-1}$ and $\beta = \left[\exp\left(0.51\sigma_R^2/(1+0.69\sigma_R^{12/5})^{5/6}\right)-1\right]^{-1}$ are parameters related to Gamma-Gamma atmospheric turbulence, where $\sigma_R^2$ is Rytov variance, and $\xi^2 = \omega_{Z_{eq}}/(2\sigma_s)$ is the ratio of the equivalent received beam radius to the standard deviation of pointing errors at the receiver [29]. $\bar{\gamma}_{FSO} = \eta^2/\sigma_{FSO}^2$ represents average SNR at the FSO receiver input.

In an urban environment, the RF transmission links spanning are subjected to multi-path fading, which can be characterized by Rayleigh distribution [24]. Accordingly, the instantaneous SNR of the RF link obeys an exponential distribution with the following probability density function (pdf) and CDF:

$$f_\gamma(\gamma) = \frac{1}{\bar{\gamma}_{RF}} e^{-\frac{\gamma}{\bar{\gamma}_{RF}}}, \quad (17)$$

$$F_\gamma(\gamma) = 1 - e^{-\frac{\gamma}{\bar{\gamma}_{RF}}}, \quad (18)$$

where $\bar{\gamma}_{RF} = 1/\sigma_{RF}^2$ is average SNR at the RF receiver input. According to (2), CDF of $\gamma_1$ random variable becomes as follows:

$$F_{\gamma_1}(\gamma) = \Pr(\max(\gamma_{1,1},\gamma_{1,2},\ldots,\gamma_{1,N}) \leq \gamma) \quad (19)$$
$$= \Pr(\gamma_{1,1} \leq \gamma, \gamma_{1,2} \leq \gamma, \ldots, \gamma_{1,N} \leq \gamma).$$

Assuming independent and identically distributed RF paths and using (18), CDF of $\gamma_1$ random variable becomes as follows:

$$F_{\gamma_1}(\gamma) = \prod_{i=1}^{N} \Pr(\gamma_{1,i} \leq \gamma) = \prod_{i=1}^{N} F_{\gamma_{1,i}}(\gamma) \quad (20)$$
$$= \left(F_{\gamma_{1,i}}(\gamma)\right)^N = \left(1 - e^{-\frac{\gamma}{\bar{\gamma}_{RF}}}\right)^N.$$

Differentiating the above equation, the pdf of $\gamma_1$ random variable becomes as follows:

$$f_{\gamma_1}(\gamma) = N\left(F_{\gamma_{1,i}}(\gamma)\right)^{N-1} f_{\gamma_{1,i}}(\gamma) \quad (21)$$
$$= \frac{N}{\bar{\gamma}_{RF}}\left(1 - e^{-\frac{\gamma}{\bar{\gamma}_{RF}}}\right)^{N-1} e^{-\frac{\gamma}{\bar{\gamma}_{RF}}}.$$

Substituting binomial expansion of $\left(1-e^{-\frac{\gamma}{\bar{\gamma}_{RF}}}\right)^{N-1}$ as $\sum_{k=0}^{N-1}\binom{N-1}{k}(-1)^k e^{-\frac{k\gamma}{\bar{\gamma}_{RF}}}$, the pdf of $\gamma_1$ random variable becomes as follows:

$$f_{\gamma_1}(\gamma) = \sum_{k=0}^{N-1}\binom{N-1}{k}(-1)^k \frac{N}{\bar{\gamma}_{RF}} e^{-\frac{(k+1)\gamma}{\bar{\gamma}_{RF}}}. \quad (22)$$

According to (15), the CDF of $\gamma_1$ random variable becomes as follows:

$$F_{\gamma_2}(\gamma) = \Pr(\max(\gamma_{FSO},\gamma_{RF}) \leq \gamma) \quad (23)$$
$$= \Pr(\gamma_{FSO} \leq \gamma, \gamma_{RF} \leq \gamma).$$

## 3 Performance of known CSI scheme

Under the idealized simplifying assumption of having perfect CSI, modulation with coherent detection require lower SNR than their non-coherent detection counterparts. However, the phase recovery error degrades the performance of the system with coherent detection, while differential detections such as DPSK are less sensitive to it. Practically, non-coherent modulations are better choices due to the carrier synchronization and the carrier recovery error. Moreover, the non-coherent detection also reduces the complexity of the receiver [24]. In this work, the BER and $P_{out}$ of DPSK and are investigated analytically.

### 3.1 Outage Probability

Given that $P_{out}(\gamma_{th}) = F_\gamma(\gamma_{th})$, and assuming that error occurs only due to the error of the access point and Base Station receivers, for the proposed detect and forward scheme, $P_{out}$ is given by [30].

$$P_{out}(\gamma_{th}) = \cup_{j=1}^{2}(\Pr\{j^{th} \text{ link is in outage}\}) = 1 - \quad (24)$$
$$\cap_{j=1}^{2}(\Pr\{j^{th} \text{ link is avalable}\}) = 1 - [1 - P_{out,1}(\gamma_{th})][1 - P_{out,2}(\gamma_{th})] = 1 - [1 - F_{\gamma_1}(\gamma_{th})][1 - F_{\gamma_2}(\gamma_{th})].$$

Assuming independent FSO and RF links, substituting (16), (18) and (20) into (24), $P_{out}$ of the proposed structure is calculated as follows:

$$P_{out}(\gamma_{th}) = 1 - \left[1 - \left(1 - e^{-\frac{\gamma_{th}}{\bar{\gamma}_{RF}}}\right)^N\right]\left[1 - \frac{\xi^2}{\Gamma(\alpha)\Gamma(\beta)}\left(1 - e^{-\frac{\gamma_{th}}{\bar{\gamma}_{RF}}}\right)G_{2,4}^{3,1}\left(\alpha\beta\kappa\sqrt{\frac{\gamma_{th}}{\bar{\gamma}_{FSO}}}\bigg|\begin{matrix}1,\xi^2+1\\ \xi^2,\alpha,\beta,0\end{matrix}\right)\right]. \quad (25)$$

Substituting binomial expansion of $\left(1-e^{-\gamma_{th}/\bar{\gamma}_{RF}}\right)^N$ in (25), $P_{out}$ of the proposed structure becomes as follows:

$$P_{out}(\gamma_{th}) = \sum_{k=0}^{N}\binom{N}{k}(-1)^k e^{-\frac{k\gamma_{th}}{\bar{\gamma}_{RF}}} + \frac{\xi^2}{\Gamma(\alpha)\Gamma(\beta)}\left(1 - e^{-\frac{\gamma_{th}}{\bar{\gamma}_{RF}}}\right)G_{2,4}^{3,1}\left(\alpha\beta\kappa\sqrt{\frac{\gamma_{th}}{\bar{\gamma}_{FSO}}}\bigg|\begin{matrix}1,\xi^2+1\\ \xi^2,\alpha,\beta,0\end{matrix}\right) - \quad (26)$$
$$\sum_{k=0}^{N}\binom{N}{k}(-1)^k \frac{\xi^2}{\Gamma(\alpha)\Gamma(\beta)}\left(e^{-\frac{k\gamma_{th}}{\bar{\gamma}_{RF}}} - e^{-\frac{(k+1)\gamma_{th}}{\bar{\gamma}_{RF}}}\right)G_{2,4}^{3,1}\left(\alpha\beta\kappa\sqrt{\frac{\gamma_{th}}{\bar{\gamma}_{FSO}}}\bigg|\begin{matrix}1,\xi^2+1\\ \xi^2,\alpha,\beta,0\end{matrix}\right).$$

Let $\xi \to 0$ at (25), then $P_{out} \to \left(1 - e^{-\frac{\gamma_{th}}{\bar{\gamma}_{RF}}}\right)^N$, that means in the proposed FSO/RF systems in spite of the common FSO systems, even when the FSO link is completely disconnected due to the effect of pointing error, the outage probability does not lead to 1, and leads to something really small and affordable, further if $N \to \infty$, then $P_{out} \to 0$, that mean the proposed multiuser system do best, even at worst case scenarios, without the use of FSO link. This is due to one of the most important properties of parallel FSO/RF links i.e. complementary physical phenomena of FSO and RF links which make connection possible even when one of links is disrupted.

### 3.2 Bit Error Rate

BER of DPSK modulation, is calculated analytically from the following equation:

$$P_e = \frac{1}{2}\int_0^\infty e^{-\gamma} F_\gamma(\gamma)d\gamma = \frac{1}{2}\int_0^\infty e^{-\gamma} P_{out}(\gamma)d\gamma. \quad (27)$$

Substituting (26) into (27) and using [28, Eq. 07.34.21.0088.01], BER of DPSK modulation becomes equal to (28), where $\varphi_1 = \left(0,\frac{1}{2},1,\frac{1+\xi^2}{2},\frac{2+\xi^2}{2}\right)$ and $\varphi_2 = \left(\frac{\xi^2}{2},\frac{1+\xi^2}{2},\frac{\alpha}{2},\frac{1+\alpha}{2},\frac{\beta}{2},\frac{1+\beta}{2},0,\frac{1}{2}\right)$.

Due to the complex structure of Meijer-G functions, it is not easy have insight about them; B.T.W. in (28) if $\xi \to 0, N = 1$, then $P_e =$

$$P_e = \frac{1}{2}\left\{\sum_{k=0}^{N}\binom{N}{k}(-1)^k \frac{1}{1+\frac{k}{\bar{\gamma}_{RF}}} + \frac{\xi^2}{\Gamma(\alpha)\Gamma(\beta)}\left(G_{5,8}^{6,3}\left(\frac{(\alpha\beta\kappa)^2}{16\bar{\gamma}_{FSO}}\bigg|\begin{array}{c}\varphi_1\\\varphi_2\end{array}\right) - \frac{1}{1+\frac{1}{\bar{\gamma}_{RF}}} G_{5,8}^{6,3}\left(\frac{(\alpha\beta\kappa)^2}{16\bar{\gamma}_{FSO}\left(1+\frac{1}{\bar{\gamma}_{RF}}\right)}\bigg|\begin{array}{c}\varphi_1\\\varphi_2\end{array}\right)\right) - \right.$$
$$\left.\sum_{k=0}^{N}\binom{N}{k}(-1)^k \frac{\xi^2}{\Gamma(\alpha)\Gamma(\beta)}\left(\frac{1}{1+\frac{k}{\bar{\gamma}_{RF}}} G_{5,8}^{6,3}\left(\frac{(\alpha\beta\kappa)^2}{16\bar{\gamma}_{FSO}\left(1+\frac{k}{\bar{\gamma}_{RF}}\right)}\bigg|\begin{array}{c}\varphi_1\\\varphi_2\end{array}\right) - \frac{1}{1+\frac{k+1}{\bar{\gamma}_{RF}}} G_{5,8}^{6,3}\left(\frac{(\alpha\beta\kappa)^2}{16\bar{\gamma}_{FSO}\left(1+\frac{k+1}{\bar{\gamma}_{RF}}\right)}\bigg|\begin{array}{c}\varphi_1\\\varphi_2\end{array}\right)\right)\right\}. \quad (28)$$

$\frac{1}{2}\frac{1}{\bar{\gamma}_{RF}+1}$ which its insight is explicit by itself; the well-known Rayleigh BER. Furthermore if one wants to have a deeper insight, he should understand that the statement $G_{5,8}^{6,3}\left(\frac{-}{f(\bar{\gamma}_{FSO},\bar{\gamma}_{RF})}\bigg|_{-}^{-}\right)$ in (28) is somehow the CDF of the system without considering the second hop RF link. It is known that the CDF is monotonically increasing. This statement is a function of the inverse of $\bar{\gamma}_{FSO}$ and $\bar{\gamma}_{RF}$; this leads to another explicit insight that increasing $\bar{\gamma}_{FSO}$ and $\bar{\gamma}_{RF}$ decreases BER in the proposed structure. Also statements of (28), each are related to $\bar{\gamma}_{FSO}$ and $\bar{\gamma}_{RF}$, but the point here is that $\bar{\gamma}_{RF}$ is more repeated in (28), which means that the proposed system is more dependent on the conditions of RF link.

## 4 Performance of unknown CSI scheme

According to (13), the CDF of $\gamma_{FSO}$ random variable is as follows:

$$F_{\gamma_{FSO}}(\gamma) = \Pr(\gamma_{FSO} \leq \gamma) = 1 - \Pr(\gamma_{FSO} \geq \gamma) \quad (29)$$
$$= 1 - \Pr\left(\frac{\gamma_1\gamma_{2,FSO}}{\gamma_{2,FSO}+C} \geq \gamma\right).$$

After mathematical simplification, (29) becomes as follows [31]:

$$F_{\gamma_{FSO}}(\gamma) = 1 - \int_0^\infty \Pr\left(\gamma_{2,FSO} \geq \frac{\gamma C}{x}\bigg|\gamma_1\right) f_{\gamma_1}(\gamma+x)dx. \quad (30)$$

Substituting (16) and (22) into (30) obtains:

$$F_{\gamma_{FSO}}(\gamma) = 1 - \sum_{k=0}^{N-1}\binom{N-1}{k}(-1)^k \frac{N}{\bar{\gamma}_{RF}} e^{-\frac{(k+1)\gamma}{\bar{\gamma}_{RF}}} \times \quad (31)$$
$$\left[\int_0^\infty e^{-\frac{(k+1)x}{\bar{\gamma}_{RF}}} dx - \frac{\xi^2}{\Gamma(\alpha)\Gamma(\beta)} \times \int_0^\infty e^{-\frac{(k+1)x}{\bar{\gamma}_{RF}}} G_{2,4}^{3,1}\left(\alpha\beta\kappa\sqrt{\frac{\gamma C}{x\bar{\gamma}_{FSO}}}\bigg|\begin{array}{c}1,\xi^2+1\\\xi^2,\alpha,\beta,0\end{array}\right) dx\right].$$

Substituting equivalent Meijer-G of $G_{2,4}^{3,1}\left(\alpha\beta\kappa\sqrt{\frac{\gamma C}{x\bar{\gamma}_{FSO}}}\bigg|\begin{array}{c}1,\xi^2+1\\\xi^2,\alpha,\beta,0\end{array}\right)$ as $G_{4,2}^{1,3}\left(\frac{1}{\alpha\beta\kappa}\sqrt{\frac{x\bar{\gamma}_{FSO}}{\gamma C}}\bigg|\begin{array}{c}1-\xi^2,1-\alpha,1-\beta,1\\0,-\xi^2\end{array}\right)$ [28, Eq. 07.34.17.0012.01], and using [28, Eq. 07.34.21.0088.01], CDF of $\gamma_{FSO}$ random variable becomes as follows:

$$F_{\gamma_{FSO}}(\gamma) = 1 - \sum_{k=0}^{N-1}\binom{N-1}{k}(-1)^k \frac{N}{k+1} \times e^{-\frac{(k+1)\gamma}{\bar{\gamma}_{RF}}}\left[1 - \frac{\xi^2 2^{\alpha+\beta-3}}{\pi\Gamma(\alpha)\Gamma(\beta)} G_{4,9}^{7,2}\left(\frac{(\alpha\beta\kappa)^2 C\gamma(k+1)}{16\bar{\gamma}_{FSO}\bar{\gamma}_{RF}}\bigg|\begin{array}{c}\psi_1\\\psi_2\end{array}\right)\right], \quad (32)$$

where $\psi_1 = \left(1,\frac{1}{2},\frac{2+\xi^2}{2},\frac{1+\xi^2}{2}\right)$ and $\psi_2 = \left(1,\frac{1+\xi^2}{2},\frac{\xi^2}{2},\frac{1+\alpha}{2},\frac{\alpha}{2},\frac{1+\beta}{2},\frac{\beta}{2},\frac{1}{2},0\right)$.
According to (14) CDF of $\gamma_{RF}$ random variable is equal to:

$$F_{\gamma_{RF}}(\gamma) = \Pr(\gamma_{RF} \leq \gamma) = 1 - \Pr(\gamma_{RF} \geq \gamma) \quad (33)$$
$$= 1 - \Pr\left(\frac{\gamma_1\gamma_{2,RF}}{\gamma_{2,RF}+C} \geq \gamma\right).$$

After mathematical simplification, the above statement comes as follows:

$$F_{\gamma_{RF}}(\gamma) = 1 - \int_0^\infty \Pr\left(\gamma_{2,RF} \geq \frac{\gamma C}{x}\bigg|\gamma_1\right) f_{\gamma_1}(\gamma+x)dx. \quad (34)$$

Substituting (18) and (22) into (34) obtains:

$$F_{\gamma_{RF}}(\gamma) = 1 - \int_0^\infty e^{-\frac{\gamma C}{x\bar{\gamma}_{RF}}}\left(\frac{N}{\bar{\gamma}_{RF}} \times \sum_{k=0}^{N-1}\binom{N-1}{k}(-1)^k e^{-\frac{(k+1)(x+\gamma)}{\bar{\gamma}_{RF}}}\right)dx. \quad (35)$$

Substituting equivalent Meijer-G of $e^{-\frac{\gamma C}{x\bar{\gamma}_{RF}}}$ as $G_{1,0}^{0,1}\left(\frac{\gamma C}{x\bar{\gamma}_{RF}}\bigg|\begin{array}{c}1\\-\end{array}\right)$ [28, Eq. 07.34.03.1081.01], (35) becomes:

$$F_{\gamma_{RF}}(\gamma) = 1 - \frac{N}{\bar{\gamma}_{RF}}\sum_{k=0}^{N-1}\binom{N-1}{k} \times \quad (36)$$
$$(-1)^k e^{-\frac{(k+1)\gamma}{\bar{\gamma}_{RF}}} \int_0^\infty e^{-\frac{(k+1)x}{\bar{\gamma}_{RF}}} G_{1,0}^{0,1}\left(\frac{\gamma C}{x\bar{\gamma}_{RF}}\bigg|\begin{array}{c}1\\-\end{array}\right) dx.$$

Using [28, Eq. 07.34.17.0012.01] and [28, Eq. 07.34.21.0088.01], CDF of $\gamma_{RF}$ random variable becomes equal to:

$$F_{\gamma_{RF}}(\gamma) = 1 - \sum_{k=0}^{N-1}\binom{N-1}{k}(-1)^k \frac{N}{k+1} \quad (37)$$
$$\times e^{-\frac{(k+1)\gamma}{\bar{\gamma}_{RF}}} G_{0,2}^{2,0}\left(\frac{\gamma C(k+1)}{\bar{\gamma}_{RF}^2}\bigg|\begin{array}{c}-\\1,0\end{array}\right).$$

### 4.1 Outage Probability

Given that $P_{out}(\gamma_{th}) = F_\gamma(\gamma_{th})$, and using (15) and assuming independent FSO and RF links, $P_{out}$ of unknown CSI scheme becomes as follows:

$$P_{out}(\gamma_{th}) = F_{\gamma_2}(\gamma_{th}) = \Pr(\max(\gamma_{FSO},\gamma_{RF}) \leq \gamma_{th}) \quad (38)$$
$$= \Pr(\gamma_{FSO} \leq \gamma_{th},\gamma_{RF} \leq \gamma_{th})$$
$$= P_{out,FSO}(\gamma_{th}) P_{out,RF}(\gamma_{th})$$

Substituting (32) and (37) into (38), and after some mathematical simplifications, $P_{out}$ of the proposed structure of CSI non-existence scheme becomes as (39).

From (38), $P_{out} = P_{out,FSO}P_{out,RF}$, which means outage disrupts this link only when both FSO and RF links go in outage. The relay of this structure is fixed gain, not adaptive, which means that even if one of the links disrupts, the system continues working without any attention; the most problem of this system is that it may also amplify noise. This is the physical reason for the mathematical complexity of (39) rather than the adaptive case of (26), because any physical phenomena affects this system and should be considered, while in (26) each part of the system were working independently due to the adaptive phenomena of the used relay. In spite of (26), even in the case of $\xi \to 0$, (39) does not give explicit insight; its main insight is that RF channel conditions are more important in system performance, because (39) is more related to RF parameters rather than FSO parameters.

$$P_{out}(\gamma_{th}) = 1 - \sum_{k=0}^{N-1}\binom{N-1}{k}(-1)^k \frac{N}{k+1}e^{-\frac{(k+1)\gamma_{th}}{\bar{\gamma}_{RF}}}\left[1 - \frac{\xi^2 2^{\alpha+\beta-3}}{\pi\Gamma(\alpha)\Gamma(\beta)}G_{4,9}^{7,2}\left(\frac{(\alpha\beta\kappa)^2 C\gamma_{th}(k+1)}{16\bar{\gamma}_{FSO}\bar{\gamma}_{RF}}\bigg|\begin{matrix}\psi_1\\\psi_2\end{matrix}\right)\right] -$$
$$\sum_{k=0}^{N-1}\binom{N-1}{k}(-1)^k \frac{N}{k+1}e^{-\frac{(k+1)\gamma_{th}}{\bar{\gamma}_{RF}}}G_{0,2}^{2,0}\left(\frac{C\gamma_{th}(k+1)}{\bar{\gamma}_{RF}^2}\bigg|\begin{matrix}-\\1,0\end{matrix}\right) +$$
$$\sum_{k=0}^{N-1}\sum_{g=0}^{N-1}\binom{N-1}{k}\binom{N-1}{g}(-1)^{k+g}\frac{N^2}{(k+1)(g+1)}e^{-\frac{(g+k+2)\gamma_{th}}{\bar{\gamma}_{RF}}}G_{0,2}^{2,0}\left(\frac{C\gamma_{th}(k+1)}{\bar{\gamma}_{RF}^2}\bigg|\begin{matrix}-\\1,0\end{matrix}\right)\left[1 - \frac{\xi^2 2^{\alpha+\beta-3}}{\pi\Gamma(\alpha)\Gamma(\beta)}G_{4,9}^{7,2}\left(\frac{(\alpha\beta\kappa)^2 C\gamma_{th}(k+1)}{16\bar{\gamma}_{FSO}\bar{\gamma}_{RF}}\bigg|\begin{matrix}\psi_1\\\psi_2\end{matrix}\right)\right]$$

(39)

### 4.2 Bit Error Rate

By insertion (39) into (27) and using [28, Eq. 07.34.21.0088.01] and [28, Eq. 07.34.21.0081.01], BER of DPSK modulation of CSI non-existence scheme becomes as (40), where $G_{p_1,q_1:p_2,q_2:p_3,q_3}^{n_1,m_1:n_2,m_2:n_3,m_3}\left(\begin{matrix}a_1,a_2,\ldots,a_{p_1}\\b_1,b_2,\ldots,b_{q_1}\end{matrix}\bigg|\begin{matrix}c_1,c_2,\ldots,c_{p_2}\\d_1,d_2,\ldots,d_{q_2}\end{matrix}\bigg|\begin{matrix}e_1,e_2,\ldots,e_{p_3}\\f_1,f_2,\ldots,f_{q_3}\end{matrix}\bigg|x,y\right)$ is the Extended Bivariate Meijer-G function [32].

It is correct that with modern computing tools, it is now easier to produce analytical expressions and these may well be horrible as is the case here. But it should be considered that the complexity of these formulations is related to the complex structure considered for the proposed system. In fact if one use modern computing tools he could not get shorter forms of these formulations; because it's well-known that Meijer-G form is the shortest possible closed-form expression one can find for any kind of mathematical expressions. And it's used as a tool in many of the well-cited papers published in FSO performance investigation.

Although the mathematical formulations of (39) and (40) are complex and do not add significant insights but it's not necessary to challenge with them, physical insights can easily be obtained by their plots. As mentioned, the main reason is that Meijer-G functions have complex structure. There are many papers in FSO system performance that used Meijer-G, without insight about mathematical expressions. Actually their aim is not to deal with mathematical formulations, they have just used math as a tool of investigating performance expressions, and brought physical insights while handling plotted figures in the results section.

The main attempt of this work is to present and investigate a new structure, as a solution of one the most challenging problems in new communication systems, i.e. consumed power by transmitter. This paper shows that at wide range of atmospheric turbulences even with the effect of pointing errors, it is possible to have a very good connection by consuming low power, without additional complexity or processing latency. It shows that there is no need to "do" implement huge coding or heavy detection as well as complicated processing or massive antennas to make communication possible. That's the point; complexity is a non-dissociable part in most of the existing communication systems, and simplicity is something forgotten. Because they should serve huge number of users with high reliability, rate and performance; this is almost not possible without complexity. But the proposed structure, only by addition of one simple FSO transreceiver, shows that it's possible. In this structure the consumer does not require to consume more power or add more complexity, and the receiver does not need to implement additional processing or complicated detection, and this is a significant practical insight of this system.

## 5 Simulation Results

In this section theoretical analysis of the proposed hybrid FSO / RF system performance are compared with MATLAB simulation results, for both known CSI and unknown CSI schemes. FSO link has Gamma-Gamma atmospheric turbulence with the effect of pointing error and RF link has Rayleigh fading. For simplicity and without loss of generality, FSO and RF links are assumed to have equal average SNR ($\bar{\gamma}_{FSO} = \bar{\gamma}_{RF} = \gamma_{avg}$). Moderate ($\alpha = 4, \beta = 1.9, \xi = 10.45$) and strong ($\alpha = 4.2, \beta = 1.4, \xi = 2.45$) regimes of Gamma-Gamma atmospheric turbulence with the effect of pointing error are investigated. Also it is assumed that $\eta = 1$ and $C = 1$.

In Fig. 2, Outage Probability of the proposed structure is plotted in terms of average SNR for moderate ($\alpha = 4, \beta = 1.9, \xi = 10.45$) and strong ($\alpha = 4.2, \beta = 1.4, \xi = 2.45$) atmospheric turbulence regimes, for both cases of known CSI and unknown CSI, number of users of $N = 2$, and outage threshold SNR of $\gamma_{th} = 10dB$. In both cases of known CSI and unknown CSI, there is little difference between system performance at moderate and strong regimes. For example, at $P_{out} = 10^{-2}$, $\gamma_{avg}$ difference is about $0.5dB$ and $2dB$ at known CSI and unknown CSI schemes, respectively. The fact that the proposed structure performs almost independent of the atmospheric turbulence intensity is one of its advantages. Because this independence, eliminates the requirement of an additional adaptive processing, that adjust system parameters to maintain performance at various atmospheric turbulence intensities. Thereby cost, power consumption, complexity, and processing latency of the proposed system are greatly reduced. Unknown CSI scheme has better performance than known CSI scheme, this is related to the constant parameter ($C$) used in unknown CSI scheme, which is adjusted manually by the operator. However this improvement in the case of unknown CSI is obtained by consuming more power in amplification block.

In Fig. 3, Outage Probability of the proposed structure is plotted in terms of average SNR for various number of users for both cases of known CSI and unknown CSI, moderate atmospheric turbulence regime ($\alpha = 4, \beta = 1.9, \xi = 10.45$) and $\gamma_{th} = 10dB$. As can be seen, in both known CSI and unknown CSI schemes, system performance is strongly dependent on number of users. Because of the independence of first-hop RF paths, the probability of availability of all received signals is equal to the product of probability of availability of individual paths. Hence, when number of users decreases, in fact number of signals used for decision at the access point decreases. Thus finding an available signal becomes easier. From this point of view, the proposed structure is power and cost effective especially within the cells with a large number of users.

$$P_e = \frac{1}{2}\Bigg\{1 - \sum_{k=0}^{N-1}\binom{N-1}{k}(-1)^k\frac{N}{k+1}\frac{1}{1+\frac{k+1}{\bar{\gamma}_{RF}}}\left[1 - \frac{\xi^2 2^{\alpha+\beta-3}}{\pi\Gamma(\alpha)\Gamma(\beta)}G_{5,9}^{7,3}\left(\frac{(\alpha\beta\kappa)^2 C(k+1)}{16\bar{\gamma}_{FSO}(\bar{\gamma}_{RF}+k+1)}\bigg|\begin{matrix}0,\psi_1\\\psi_2\end{matrix}\right)\right] -$$
$$\sum_{k=0}^{N-1}\binom{N-1}{k}(-1)^k\frac{N}{k+1}\frac{1}{1+\frac{k+1}{\bar{\gamma}_{RF}}}G_{1,2}^{2,1}\left(\frac{C(k+1)}{\bar{\gamma}_{RF}(\bar{\gamma}_{RF}+k+1)}\bigg|\begin{matrix}0\\1,0\end{matrix}\right) +$$
$$\sum_{k=0}^{N-1}\sum_{g=0}^{N-1}\binom{N-1}{k}\binom{N-1}{g}(-1)^{k+g}\frac{N^2}{(k+1)(g+1)}\frac{1}{1+\frac{g+k+2}{\bar{\gamma}_{RF}}}\Bigg[G_{1,2}^{2,1}\left(\frac{C(k+1)}{\bar{\gamma}_{RF}(\bar{\gamma}_{RF}+k+g+2)}\bigg|\begin{matrix}0\\1,0\end{matrix}\right) -$$
$$\frac{\xi^2 2^{\alpha+\beta-3}}{\pi\Gamma(\alpha)\Gamma(\beta)}G_{1,0:0,2:4,9}^{1,0:2,0:7,2}\left(\begin{matrix}1\\-\end{matrix}\bigg|\begin{matrix}-\\0,1\end{matrix}\bigg|\begin{matrix}\psi_1\\\psi_2\end{matrix}\bigg|\frac{C(k+1)}{\bar{\gamma}_{RF}(\bar{\gamma}_{RF}+k+g+2)},\frac{(\alpha\beta\kappa)^2 C(g+1)}{16\bar{\gamma}_{FSO}(\bar{\gamma}_{RF}+k+g+2)}\right)\Bigg]\Bigg\}$$

(40)

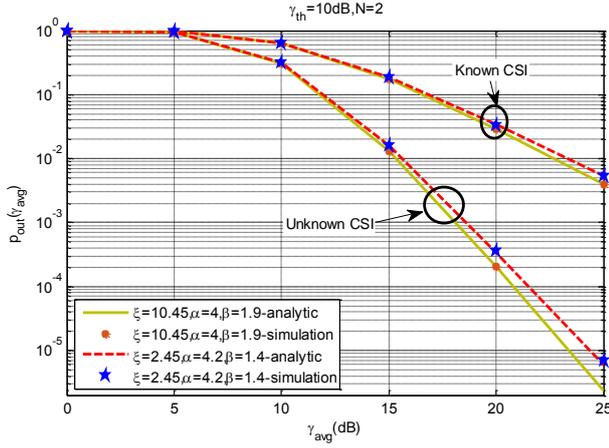

**Fig. 2**: Outage Probability of the proposed structure in terms of average SNR for moderate ($\alpha = 4, \beta = 1.9, \xi = 10.45$) and strong ($\alpha = 4.2, \beta = 1.4, \xi = 2.45$) atmospheric turbulence regimes, for both cases of known CSI and unknown CSI, number of users of $N = 2$ and $\gamma_{th} = 10dB$.

Because without using additional processing or complexity, system performance becomes favourable even at low SNRs.

Dependence on the number of users in the case of known CSI is less than the unknown CSI. Because when CSI is available, access point detects the received signal based on an adaptive threshold and adjust itself to the conditions. But amplification by a fixed gain, increases the performance difference between different numbers of users. Because noise and fading coefficients are also amplified.

In Fig. 4, Bit Error Rate of the proposed structure is plotted in terms of average SNR for various numbers of users, for both cases of known CSI and unknown CSI, and moderate atmospheric turbulence regime ($\alpha = 4, \beta = 1.9, \xi = 10.45$). It can be seen that at $\gamma_{avg} \leq 5dB$ known CSI scheme performs better than unknown CSI, but at $\gamma_{avg} \geq 5dB$, unknown CSI scheme has better performance. This is related to the constant parameter ($C$) in unknown CSI scheme, which is assumed to be unit ($C = 1$). If $C$ was chosen smaller the unknown CSI scheme would show better performance at all $\gamma_{avg}$.

It can be seen that at low $\gamma_{avg}$, performance of both known CSI and unknown CSI schemes depends on the number of users within the cell, but at high $\gamma_{avg}$, this dependence decreases in the case of known CSI and increases in the case of unknown CSI.

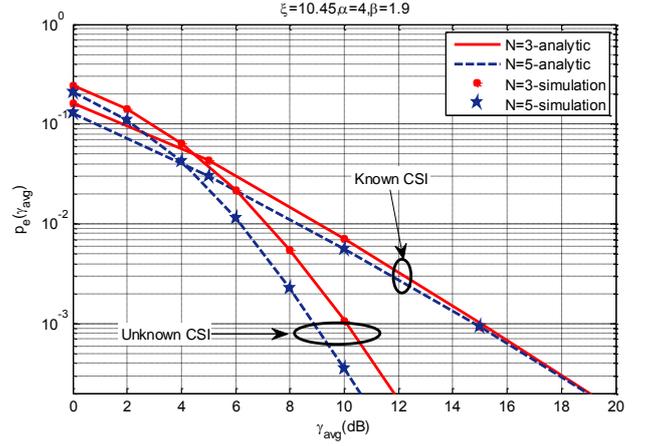

**Fig. 4**: Bit Error Rate of the proposed structure in terms of average SNR for various numbers of users, for both cases of known CSI and unknown CSI and moderate atmospheric turbulence regime ($\alpha = 4, \beta = 1.9, \xi = 10.45$).

In Fig. 5, Bit Error Rate of the proposed structure is plotted in terms of average SNR for moderate ($\alpha = 4, \beta = 1.9, \xi = 10.45$) and strong ($\alpha = 4.2, \beta = 1.4, \xi = 2.45$) regimes of Gamma-Gamma atmospheric turbulence with the effect of pointing error for both cases of known CSI and unknown CSI, and number of users of $N = 2$. As can be seen, there is little performance difference between moderate and strong atmospheric turbulence regimes. Therefore, the proposed system does not require adaptive processing or consuming much more power in order to maintain its performance. This is important especially for cells which encounter frequent changes of atmospheric turbulence intensity during the day. In these areas, frequent adaption of system parameters does not work, it may also cause more performance degrade. But in the proposed system performance maintains without additional processing or power consumption.

As can be seen, the main advantage of the proposed structure is its favourable performance even at low SNRs. This fact makes it suitable for power demand applications such as mobile communication systems in which a small mobile battery supplies transmitter power.

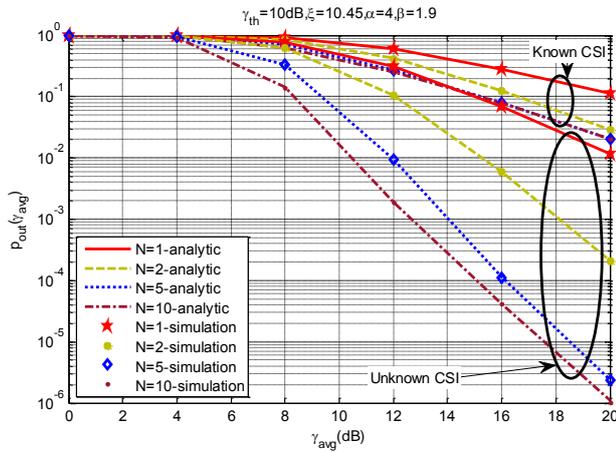

**Fig. 3**: Outage Probability of the proposed structure in terms of average SNR for various number of users, for both cases of known CSI and unknown CSI, moderate atmospheric turbulence regime ($\alpha = 4, \beta = 1.9, \xi = 10.45$) and $\gamma_{th} = 10dB$.

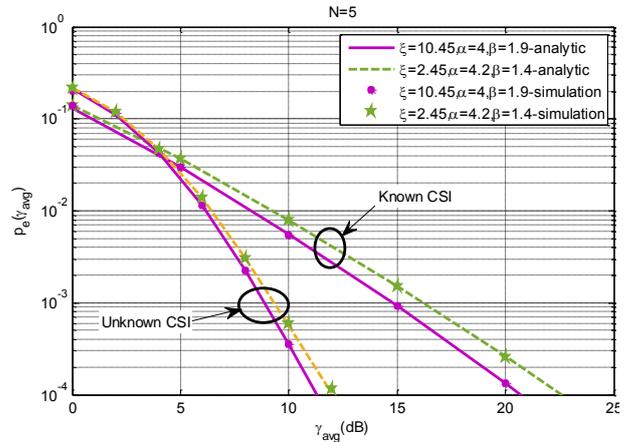

**Fig. 5**: Bit error rate of the proposed structure in terms of average SNR for moderate ($\alpha = 4, \beta = 1.9, \xi = 10.45$) and strong ($\alpha = 4.2, \beta = 1.4, \xi = 2.45$) regimes of Gamma-Gamma atmospheric turbulence with the effect of pointing error for both cases of known CSI and unknown CSI, and number of users of $N = 2$.

# 6 Conclusion

In this paper, a novel model is presented for hybrid FSO / RF communication system, in which an access point connects users within a building to the Base Station via a hybrid parallel FSO / RF link. FSO link has Gamma-Gamma atmospheric turbulence with the effect of pointing error and RF link has Rayleigh fading. Performance of the proposed system is investigated for both cases of known CSI and unknown CSI at the access point. For the first time, closed-form expressions are derived for BER and $P_{out}$ of the proposed structure. MATLAB simulations verified the derived expressions.

In this work for the first time the effect of the number of users within the building on the performance of a dual-hop hybrid FSO /RF link is investigated at moderate to strong atmospheric turbulences. Performance of the system improves by increasing number of users, because the access point selects the user with the highest SNR. However, in the case of unknown CSI this improvement is more. When CSI is available, system parameters can get adapted to the conditions, thus the dependence to the number of users is low. But unknown CSI scheme is more dependent, because the selected signal is amplified by a fixed gain; the higher this gain the bigger performance difference between various numbers of users. Communication systems which encounter fast and frequent changes in atmospheric turbulence intensity, need independent performance of atmospheric turbulence intensity. It is shown that there is little performance difference between moderate and strong atmospheric turbulence regimes. The proposed system maintains performance without additional processing or power consumption and complexity, also it offers favourable performance even at low SNRs. Therefore, it is economically affordable and particularly suitable for power demanded applications such as multiuser mobile communication systems.